\documentstyle[12pt,aaspp4,flushrt]{article}

\slugcomment{To appear in the Astrophysical Journal Letters}

\begin{document}

\title{THE ORIGIN OF THE SOLAR FLARE WAITING-TIME DISTRIBUTION}

\author{M.S. Wheatland}
\affil{Research Centre for Theoretical Astrophysics, School of Physics,
University of Sydney, NSW 2006, Australia}

\begin{abstract}
It was recently pointed out that the distribution of times between
solar flares (the flare waiting-time distribution) follows a power law, 
for long waiting times. Based on 25 years of soft X-ray flares observed 
by Geostationary Operational Environmental Satellite (GOES) instruments 
it is shown that 1.\ the waiting-time distribution of flares is consistent 
with a time-dependent Poisson process, and 2.\
the fraction of time the Sun spends with different flaring rates 
approximately follows an exponential distribution. The second result is a 
new phenomenological law for flares. It is shown analytically how the 
observed power-law behavior of the waiting times originates in the 
exponential distribution of flaring rates. These results are argued to be 
consistent with a non-stationary avalanche model for flares. 
\end{abstract}
\keywords{MHD -- Sun: activity -- Sun: corona -- Sun: flares -- Sun: X-rays}

\section{Introduction}

\noindent
The distribution of times between flares (``waiting times'') gives
information about whether flares occur as independent events, and also
provides a test for models for flare statistics. For example, the avalanche
model for flares (\cite{lu&ham91}, \cite{lu&93}) is a model designed to 
reproduce the observed power-law distributions of flare energy and 
duration. Flares are described as redistribution events in a cellular 
automaton (CA) that is driven to a self-organized critical state. Because 
the system is driven at a constant (mean) rate and flares occur as 
independent events, the model makes the specific prediction that the flare 
waiting-time distribution (WTD) is a simple exponential, consistent with a 
Poisson process. 

Observational determinations of the flare WTD have given varying results.
Determinations based on hard X-ray observations have focused on the
distribution of short waiting times (seconds -- hours). Biesecker
(1994) found the WTD for hard X-ray bursts observed by the Burst and 
Transient Source Experiment on the Compton Gamma Ray Observatory to be
consistent with a time-dependent Poisson process, i.e.\
one in which the mean flaring rate is time-varying. This result is 
consistent with a non-stationary avalanche model for flares (an 
avalanche model driven with a non-constant rate). However, Wheatland et 
al.\ (1998) found an overabundance of short waiting times (by comparison 
with a time-dependent Poisson process) in hard X-ray bursts observed by 
the Interplanetary Cometary Observer (ICE) spacecraft. (For another 
determination of the WTD based on hard X-ray, see \cite{pea&93}.)

Recently, the distribution of times between soft X-ray flares observed by
the Geostationary Operational Environmental Satellite sensors (GOES) between
1976 and 1996 was examined by Boffeta et al.\ (1999). The advantage of the 
GOES data is that it provides a long sequence of data with few gaps, and 
so the flare 
WTD can be examined for long waiting times. Boffeta et al.\ found that the 
distribution follows a power law for waiting times greater than a few hours. 
They argued that this result is inconsistent with the avalanche model, and 
that the appearance of a power law suggests a turbulence model for the 
origin of flares. 

In this paper the GOES data is re-examined. It is shown that the observed,
power-law like WTD is consistent with a piecewise-constant Poisson process, 
and hence with the non-stationary avalanche model. Further, it is shown that
the time distribution of rates of the GOES flares averaged over several solar 
cycles is approximately exponential. This is a new 
phenomenological law for flaring. Finally, it is shown analytically how a 
piecewise-constant Poisson process with an exponential distribution of 
rates has a WTD that is power-law distributed for long waiting times, 
consistent with the observations.

\section{Data analysis}

\noindent
The data examined here is the catalog of flares observed during 1975-1999 
by the 1-8\,\AA ~GOES sensors (see \cite{gar94} for details of the GOES 
instrument). The chosen period of time covers three solar cycles 
(21, 22 and 23). Because the soft X-ray background rises with the solar 
cycle, flares are undercounted near solar maximum, by comparison with
solar minimum. Hence only those flares with a peak flux greater than a 
threshold value ($10^{-6}\,{\rm W m}^{-2}$, corresponding to a GOES C1.0 class
flare) are included in the study. This leaves a total of 32,563 flares.

Figure~1 shows the WTD for the GOES events included in this study
(histogram), constructed from differences between start times of flares. 
The figure agrees well with Figure~1 in Boffeta et al.\ (1999),
and in particular shows the same power-law like behavior for waiting times 
greater than a few hours. The index of the power law is about 
$-2.16\pm 0.05$ (for waiting times greater than 10 hours), which may be 
compared with Boffeta et al.'s estimate of $-2.4\pm 0.1$. The difference 
in power-law indices is due to the restriction to flares greater than
class C1.0 -- Boffeta et al.\ included all flares in their determination
of the waiting-time distribution. Error bars are plotted on the histogram
in Figure~1, corresponding to the square root of the number of waiting 
times in each bin. The meaning of the solid and dashed curves in the figure 
is explained below.

To compare the observed occurrence of flares with that expected from a
time-dependent Poisson process, it is necessary to determine the mean rate
of flaring as a function of time. For this step a Bayesian procedure
devised by Scargle (1998) was used. (The same procedure was used in Wheatland
et al.\ 1998.) The method takes a sequence of times of events and 
determines a decomposition into intervals of time when the observed event
occurrence is consistent with a (constant rate) Poisson process. These
intervals are characterized by a duration $t_i$ and a rate $\lambda_i$, and
are referred to as ``Bayesian blocks.'' The procedure has only one free
parameter, a ``prior odds ratio,'' which disfavors further segmentation of
intervals when the single rate and dual-rate Poisson model are almost equally
likely (see \cite{sca98} for further details). However, very similar
results are achieved for different choices of this ratio. In the following
analysis, the value ${\tt PRIOR\_ODDS=2}$ is used. 

Figure~2 shows the results of the application of the Bayesian procedure to
the GOES data. The method has decomposed the 25 years of flaring into 
390 Bayesian blocks. The rate of flaring 
is observed to vary with the solar cycle, as expected, and also exhibits
short time-scale variations. Relatively long intervals with a constant
rate are also observed.

A Poisson process with a constant rate $\lambda$ has a WTD given by
$P(\Delta t)=\lambda\exp(-\lambda\Delta t)$, where $\Delta t$ describes a
waiting time. The WTD distribution for a 
piecewise-constant Poisson process with rates $\lambda_i$ and intervals $t_i$
may be approximated by 
\begin{equation}\label{eq:tdpp}
P(\Delta t)\approx\sum_i\varphi_i\lambda_i \exp(-\lambda_i\Delta t),
\end{equation}
where 
\begin{equation}
\varphi_i=\frac{\lambda_it_i}{\sum_j \lambda_j t_j}
\end{equation}
is the fraction of events associated with a given rate $\lambda_i$.

The rates and intervals shown in Figure~2 were used to construct a model 
WTD, from (\ref{eq:tdpp}). The result is plotted in Figure~1 as the solid 
curve. It is clear that there is good qualitative agreement between the
observed and model WTDs. In particular, the model distribution reproduces
power-law like behavior for long waiting times, and is relatively constant
for short waiting times. There is some discrepancy between the curves,
e.g.\ there are too few observed short waiting times, and too many observed
waiting times near the rollover in the distribution. However, it is likely
that there are errors in the observational determination of the WTD. For
example, short waiting times are likely to be missed due to the overlap of
flares close in time. Also, the Bayesian method for determining rates from
the data may produce some erroneous rates and intervals, and the 
expression~(\ref{eq:tdpp}) and the
decomposition into a piecewise constant Poisson process involve
approximations that are difficult to precisely quantify. The good 
qualitative agreement of the model and observed distributions is taken as 
strong evidence that the GOES flares occur as a time-varying Poisson process.  

\section{The origin of the power-law behavior}

\noindent
The Bayesian procedure decomposed the GOES time series into a large
number of Bayesian blocks. For a piecewise-constant Poisson process 
involving a large number of rates, the summation in~(\ref{eq:tdpp}) may be 
replaced by an integral:
\begin{equation}\label{eq:lap}
P(\Delta t)=\frac{1}{\lambda_0}\int_{0}^{\infty}f(\lambda)\lambda^2 
  e^{-\lambda\Delta t}\,d\lambda,
\end{equation}
where $f(\lambda)\,d\lambda$ is the fraction of time that the flaring rate
is in the range $(\lambda,\lambda+d\lambda)$, and
\begin{equation}
\lambda_0=\int_{0}^{\infty}\lambda f(\lambda)\,d\lambda
\end{equation}
is the mean rate of flaring.

The expression~(\ref{eq:lap}) for the WTD of a piecewise-constant Poisson
process depends only on the time distribution of the rates of flaring,
$f(\lambda)$. Figure~3 shows this distribution (the histogram), constructed 
from the rates and intervals shown in Figure~2. This figure reveals the 
remarkable 
fact that the rate of flaring  -- effectively averaged over several solar 
cycles -- follows an exponential distribution, a result that
does not appear to have been noted in the literature before. 
The observed distribution may be approximated by
\begin{equation}\label{eq:exp}
f(\lambda)=\lambda_0^{-1}\exp(-\lambda/\lambda_0), 
\end{equation}
where $\lambda_0\approx 0.15\,{\rm hour}^{-1}$ is obtained from the total
number of flares divided by the total observing time. 
Equation~(\ref{eq:exp}) is shown by the straight line in Figure~3. The 
observed distribution does not agree exactly with the exponential form. 
The cumulative probability distribution corresponding to Figure~3 (this 
distribution is preferable to 
the differential distribution because it involves no binning) was compared 
with the model distribution corresponding to (\ref{eq:exp}) using the 
Kolmogorov-Smirnov test (\cite{bab&fei96}). 
This test excludes the possibility that the two distributions are the same
at a high level of significance. However, the observationally-inferred
distribution of rates is somewhat uncertain, and the exponential model
clearly provides a good first approximation to the observed distribution.

Substituting~(\ref{eq:exp}) into~(\ref{eq:lap}), the integral may be
evaluated to give
\begin{equation}\label{eq:result}
P(\Delta t)=\frac{2\lambda_0}{(1+\lambda_0\Delta t)^3}.
\end{equation}
Equation~(\ref{eq:result}) is plotted in Figure~1 as the dashed
curve. For short waiting times ($\Delta t\ll \lambda_0^{-1}$), 
equation~(\ref{eq:result}) approaches the value $P(\Delta t)=2\lambda_0$. 
For long waiting times ($\Delta t\gg \lambda_0^{-1}$), the distribution has 
the power-law form $P(\Delta t)\sim 2\lambda_0^{-2}(\Delta t)^{-3}$. 
Hence equation~(\ref{eq:result}) accounts for the qualitative behavior 
of the WTD, in particular the power-law behavior for large waiting times,
the location of the rollover to power-law
form ($\Delta t\approx \lambda_0^{-1}$), and the approximate index of the 
power-law tail. The behavior of the observed WTD is seen to 
originate from a time-dependent Poisson process with an approximately 
exponential distribution of rates.

\section{Discussion}
\noindent
In this paper the waiting-time distribution for 25 years of GOES soft 
X-ray flares (of greater than C1.0 class) has been investigated. The observed 
WTD is found to be qualitatively consistent with a piecewise-constant 
Poisson process, with a time history of rates determined from the data 
using a Bayesian procedure. This result indicates that the GOES flares are 
independent, random events. There does not appear to be good evidence in 
the GOES events for flare sympathy, or for long-term correlations in the
times of flare occurrence.

The GOES WTD displays a power-law tail for long waiting times, as pointed 
out by Boffeta et al.\ (1999), and confirmed here. In this paper the 
power-law behavior is demonstrated to originate from two basic
assumptions, that are well supported by the data: 1.\ that flare process 
is Poisson, and 2.\ that the distribution of flaring rates follows an
approximate exponential. Subject only to these assumptions, the 
theoretical WTD is equation~(\ref{eq:result}), which reproduces the 
qualitative features of the observed WTD, including the power-law tail. 
There is some discrepancy 
between equation~(\ref{eq:exp}) and the observationally determined WTD.
For example, the observational determination of the power-law index of the
tail of the distribution is around $-2.2$ (Boffeta et al.\ found 
$-2.4\pm 0.1$), whereas equation~(\ref{eq:result}) predicts an index of 
$-3$. This difference is most likely due to the departure of the observed
distribution of flaring rates from a simple exponential form, particularly
for low flaring rates (which influence the behavior of the WTD for long
waiting times). 

This paper presents the new result that the probability of flare occurrence
per unit time, when averaged over the solar cycle, follows an approximate 
exponential distribution (see Figure~3). This is a new phenomenological 
law for flaring, that must be explained by any theory for the origin of 
flare energy. The rate of flare occurrence reflects the total rate of 
energy release in flaring, which must match the rate of energy supply to 
the system. Hence it follows that the rate at which energy is supplied 
to the corona for flaring also follows an exponential distribution. It 
is clear from Figure~2 that this new law does not hold 
instantaneously -- e.g.\ at times of maxima of the cycle, there are few low 
flaring rates. For certain periods of time during each solar cycle the 
flaring rate is approximately constant. From these points it also follows 
that the observed flare WTD is time-dependent, and may have a different
form depending upon the interval of observation. If the WTD is constructed 
for a short period of observation, during which time the rate of flaring 
is approximately constant, then the distribution will resemble an 
exponential. The power-law tail of the WTD appears in the GOES data taken 
over several solar cycles, during which time there is wide variation in 
the flaring rate. For shorter periods of observation the power-law form 
might not appear, depending on whether there is sufficient variation in the 
flaring rate. The time-dependence and cycle-dependence of the rate and 
waiting-time distributions will be investigated in more detail in future work. 

In this paper, waiting times between flares from all active regions 
present on the Sun have been considered, so that the Sun is treated as a
single flaring system. Boffeta et al.\ (1999) also considered flares in
individual active regions, as identified (in the GOES catalog) from 
H$\alpha$ events. The distribution for waiting times in individual active
regions was found to be similar to that from all active regions. In future 
work the WTD in individual active regions will be considered in more detail.  

The results presented in this paper are consistent with the avalanche model
for flares. Although avalanche cellular automata produce an exponential 
WTD when driven with a constant rate, if the rate of driving is varied so 
that the distribution of rates is exponential, then the resulting 
model (referred to here as a non-stationary avalanche model) should 
reproduce the qualitative features of the observed WTD. There is no need 
to consider models that produce a power-law WTD through long-term 
correlations between events
(e.g.\ models of MHD turbulence, cf.\ Boffeta et al.\ 1999), because the 
WTD is seen to be a simple consequence of the statistics of independent 
flare events together with an exponential distribution of flaring rates. 

The author acknowledges the support of a U2000 Post-doctoral Fellowship at
the University of Sydney.

\newpage

\begin{figure}[H]
\plotone{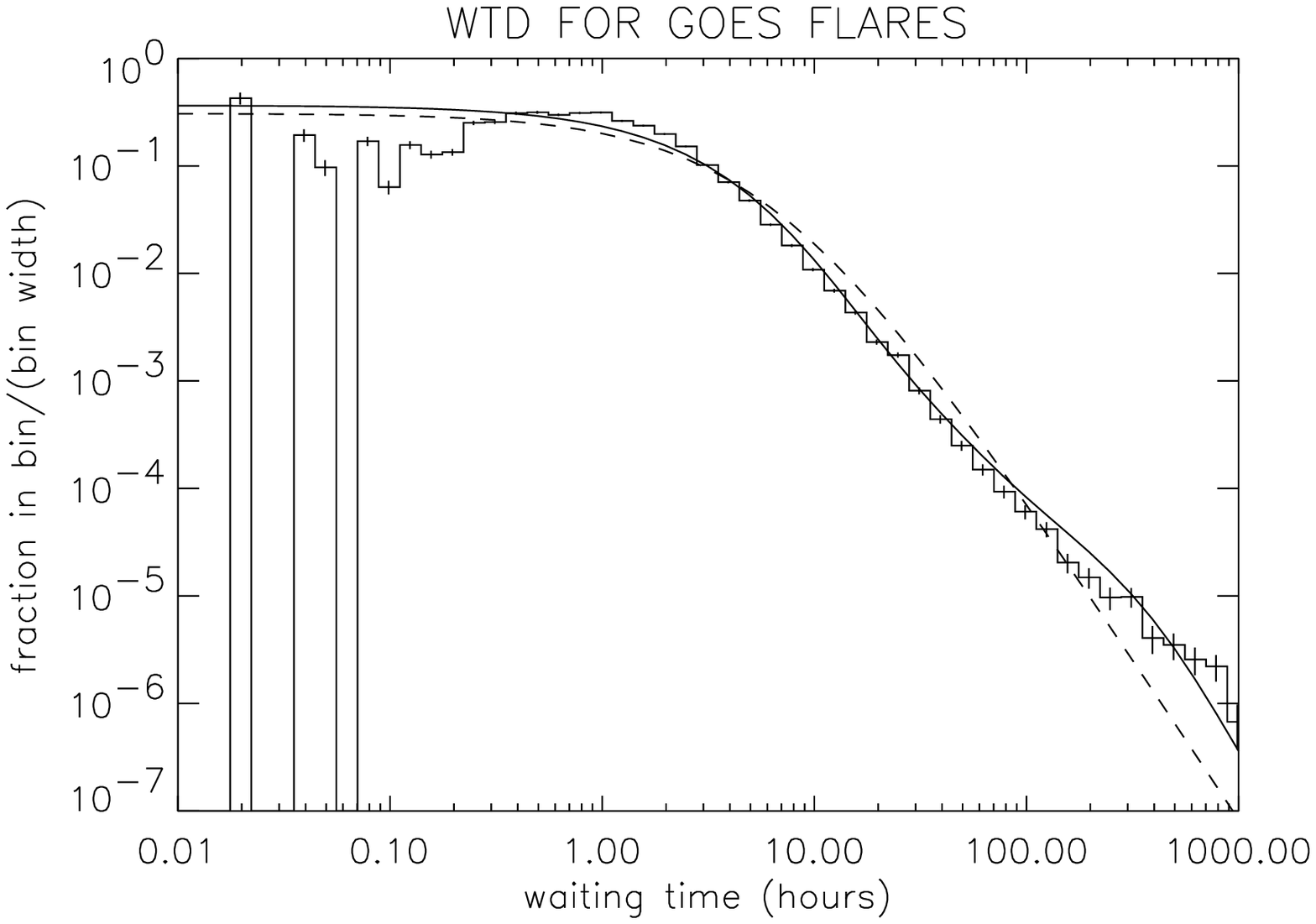}
\caption{
The WTD for the GOES flares (histogram). The solid curve
shows the result for a time-dependent Poisson process with a distribution
of rates estimated from the data. The dashed curve is a theoretical
distribution corresponding to independent events with an exponential
distribution of rates. 
}
\end{figure}

\newpage

\begin{figure}[H]
\plotone{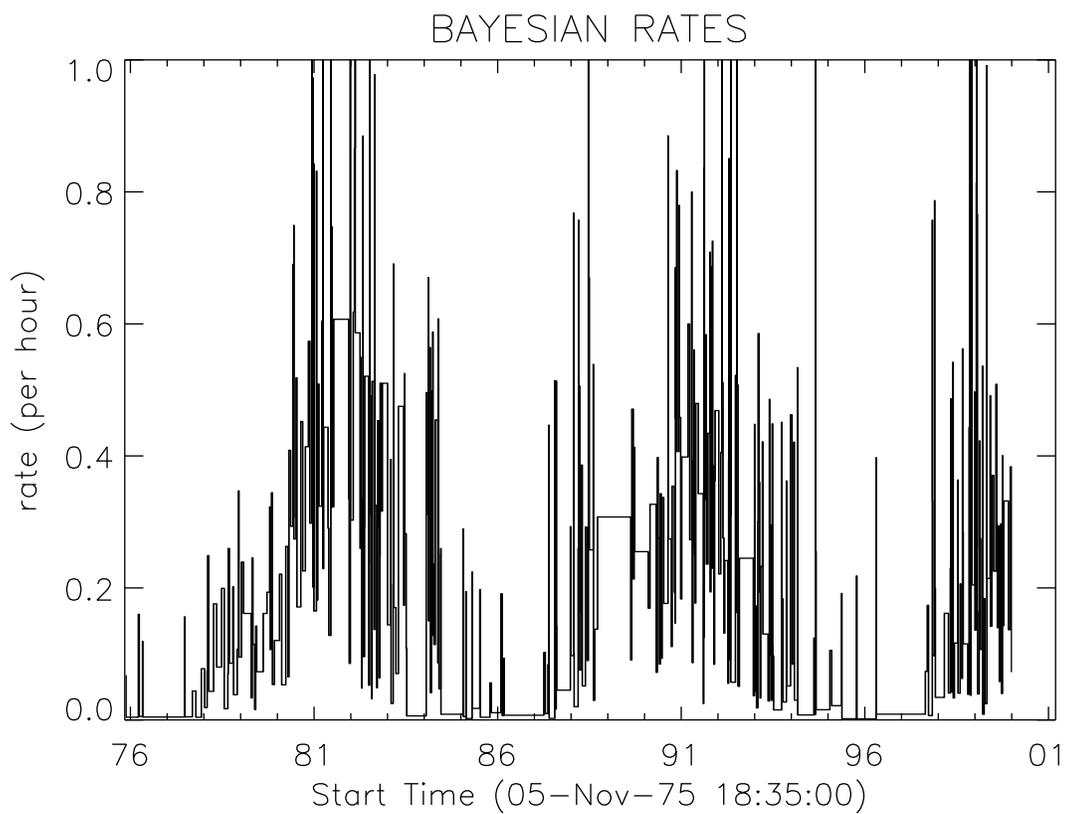}
\caption{ The ``Bayesian blocks'' decomposition of the rate of occurrence 
of the GOES flares.
}
\end{figure}

\newpage

\begin{figure}[H]
\plotone{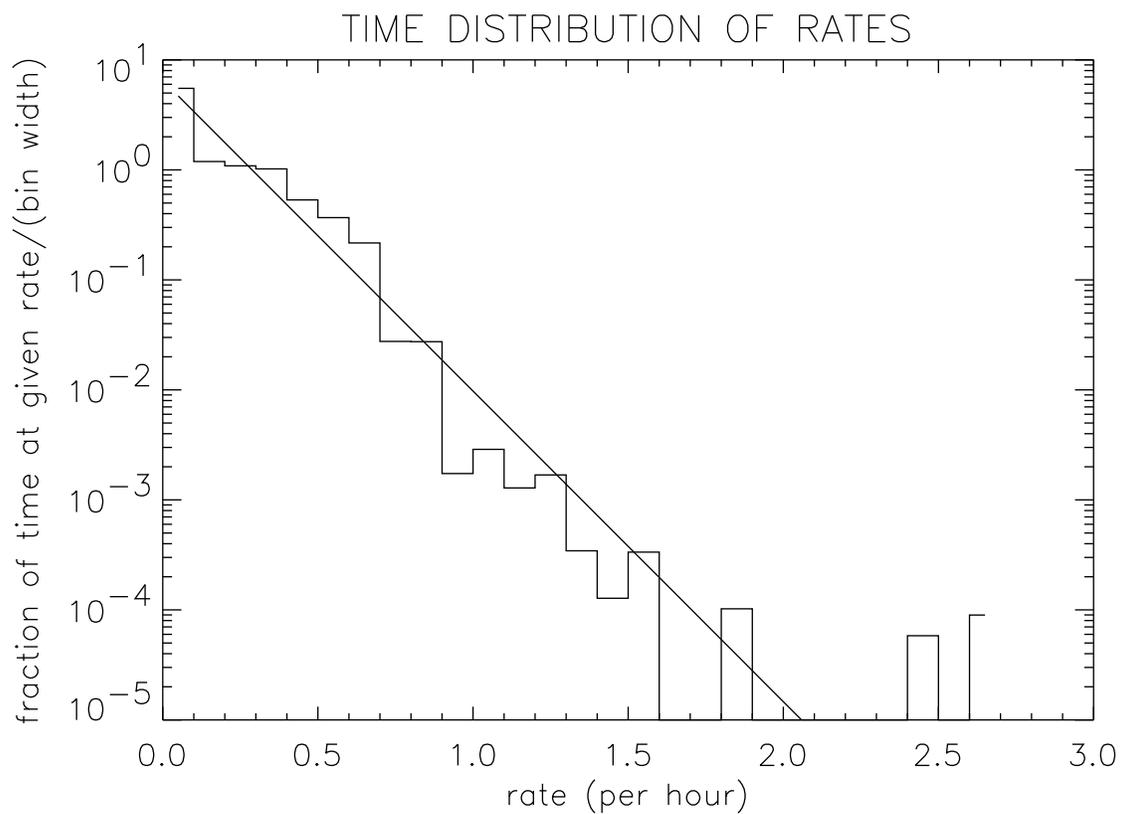}
\caption{
The distribution of flaring rates, based on the Bayesian
rate estimates of Figure~2.
}
\end{figure}

\end{document}